\documentclass{DISproc}

\usepackage{times,fancyhdr}
\usepackage{cite}
\usepackage{graphicx}
\usepackage{subfig}

\begin{document}

\newcommand{\dif}{{\rm{d}}}
\newcommand{\e}{\rm{e}}
\newcommand{\B}{\mathcal{B}}
\newcommand{\R}{\mathcal{R}}
\newcommand{\K}{\mathcal{K}}
\newcommand{\A}{\mathcal{A}}
\newcommand{\be}{\begin{equation}}
\newcommand{\ee}{\end{equation}}
\newcommand{\bi}{\begin{itemize}}
\newcommand{\ei}{\end{itemize}}
\newcommand{\med}{\medskip \\}
\newcommand{\bq}{\mathbf{q}}
\newcommand{\bk}{\mathbf{k}}
\newcommand{\bl}{\mathbf{l}}
\newcommand{\non}{\nonumber\\}
\newcommand{\asbar}{\bar{\alpha}_s}

\title{The non-forward BFKL equation and infrared effects}

\author{{\slshape G. Chachamis$^1$, A. Sabio Vera$^2$, C. Salas$^2$}\\[1ex]
$^1$Paul Scherrer Institut, CH-5232 Villigen PSI, Switzerland\\
$^2$Instituto de F{\' \i}sica Te{\' o}rica UAM/CSIC, Nicol{\'a}s Cabrera 15, UAM, E-28049 Madrid, Spain}

\contribID{214}

\doi  

\maketitle

\begin{abstract}
An iterative solution best suited for a Monte Carlo implementation is presented for the non-forward BFKL equation in a generic color representation. We introduce running coupling effects compatible with bootstrap to all orders in perturbation theory. A numerical analysis is given showing a smooth transition from a hard to a soft pomeron when accounting for running effects.
\end{abstract}

\section{Introduction}

The solution to the LL BFKL equation projected in the color singlet in the non-forward case was first calculated in the seventies by Balitski, Fadin, Kuraev and Lipatov~\cite{Fadin:1975cb, Lipatov:1976zz, Kuraev:1976ge, Kuraev:1977fs,Balitsky:1978ic}. Three decades later a solution based on a sum of iterations of the kernel in transverse momentum space was proposed at LL~\cite{Andersen:2004tt} and Next to Leading Logarithmic(NLL)~\cite{Andersen:2003wy} accuracy and used to build up a Monte Carlo code to provide numerical studies of it. Similar studies for the color octet representation were given in~\cite{Kwiecinski:1996fm,Schmidt:1996fg,Orr:1997im}.

The present work is based on the formalism given in~\cite{Andersen:2004tt}. The study is extended to give a solution in a general color group representation $\cal{R}$ and presented in a way such that all the infrared divergences appear as an overall factor in the gluon Green function so that a finite piece can be identified and treated numerically. In order to account for higher order corrections and study the properties of the infrared the running of the coupling is introduced. How to account for it is a non well-defined problem. There is no theoretical strong restriction on it and different possibilities have been suggested in the literature~\cite{Fadin:1998py,Kovchegov:1998ae, Thorne:1999rb,Braun:1994mw,Levin:1994di,Kovchegov:2006vj}. We insert it in a way consistent with gluon reggeization, as proposed in~\cite{Braun:1994mw, Levin:1994di,Kovchegov:2006vj}, which naturally leads to the appearance of renormalon singularities in the infrared. The fact that the solution is given in transverse momentum and rapidity space makes possible to study diffusion properties, analyzed in Sec.~\ref{sec:diffusion}. The aim of this review is to give a very short qualitative explanation of the main points of the work presented. We refer the reader to the article in preparation~\cite{eta} and the references given here for the calculation.

\section{Non-forward BFKL equation in a generic color representation}

The infrared divergences that appear in a general color representation can be written as an overall factor in the gluon Green 
function. To show this one has to regularize half of the divergences in the gluon Regge trajectory using dimensional regularization ($D=4-2 \,\epsilon$) and introduce a mass parameter $\lambda$ for the remaining ones. In doing it, the dependence on $\lambda$ cancels out with the contribution of the real emissions while the one on $\epsilon$ remains in the factorized term, leading to a solution to the non-forward BFKL equation independent of $\lambda$ for $\lambda \rightarrow 0$. 

The divergent term depends on the choice of the color projection. In the singlet representation, for example, $c_1 = 1$ and all infrared divergences cancel out. The non-forward equation for the color singlet can be solved in an iterative way by going back to momentum space using a Mellin transform. Details on how this is down are given in~\cite{Andersen:2003wy}. The result obtained is the product of an exponential term depending on the $\lambda$ and $1/\epsilon$ parameters and a finite part that we denote by ${\cal H} \left({\bf q}_1,{\bf q}_2;{\bf q};{\rm Y}\right)$. We analyze this function in sec.~\ref{sec:numerics}. It is convenient for the numerical study to express the gluon Green function ${\cal F} (\bq_1, \bq_2; Y)$ as a function of the azimuthal angle between the two-dimensional vectors ${\bf q}_1$ and ${\bf q}_2$, its Fourier conjugate variable or conformal spin $n$ and the anomalous dimension $\gamma$. 

\subsection{Infrared effects}

A way to introduce running coupling effects at in the analysis of a $2\to 2$ partonic process at LL accuracy is to replace the reggeized gluon propagating in the t-channel by a gluon (or renormalon) chain~\cite{Zakharov:1992bx}. Our choice of accounting for the running is based on this approach. A comparison of the LL gluon trajectory with the new one tells us that the only needed change to be done in the analytic expressions is the replacement $\bk^2 \to \eta(\bk)$, with $\eta(\bk)\equiv \bk^2/\asbar(\bk^2)$. In order to define the new BFKL kernel the bootstrap condition is imposed so that gluon reggeization is still justified. This procedure naturally leads to the appearance of renormalons as power corrections that could let us learn about the properties of the infrared. Details on this setup can be found in~\cite{Braun:1994mw, Levin:1994di,Kovchegov:2006vj}.

Concerning the choice for the running we use a parametrization which freezes in the infrared and it is consistent with global data of infrared power corrections to perturbative observables~\cite{Webber:1998um}. When the external transverse momentum scales in the gluon Green function are perturbative enough, this model for the running cannot be distinguished with a perturbative one with a Landau pole. Nonetheless, we do find sensitivity to the IR finite model for sufficient small values of these scales.

\section{Numerical analysis for the color singlet}\label{sec:numerics}

\begin{wrapfigure}{r}{0.45\textwidth}
\vspace{-.7cm}
  \centering
  \includegraphics[width=.45\textwidth]{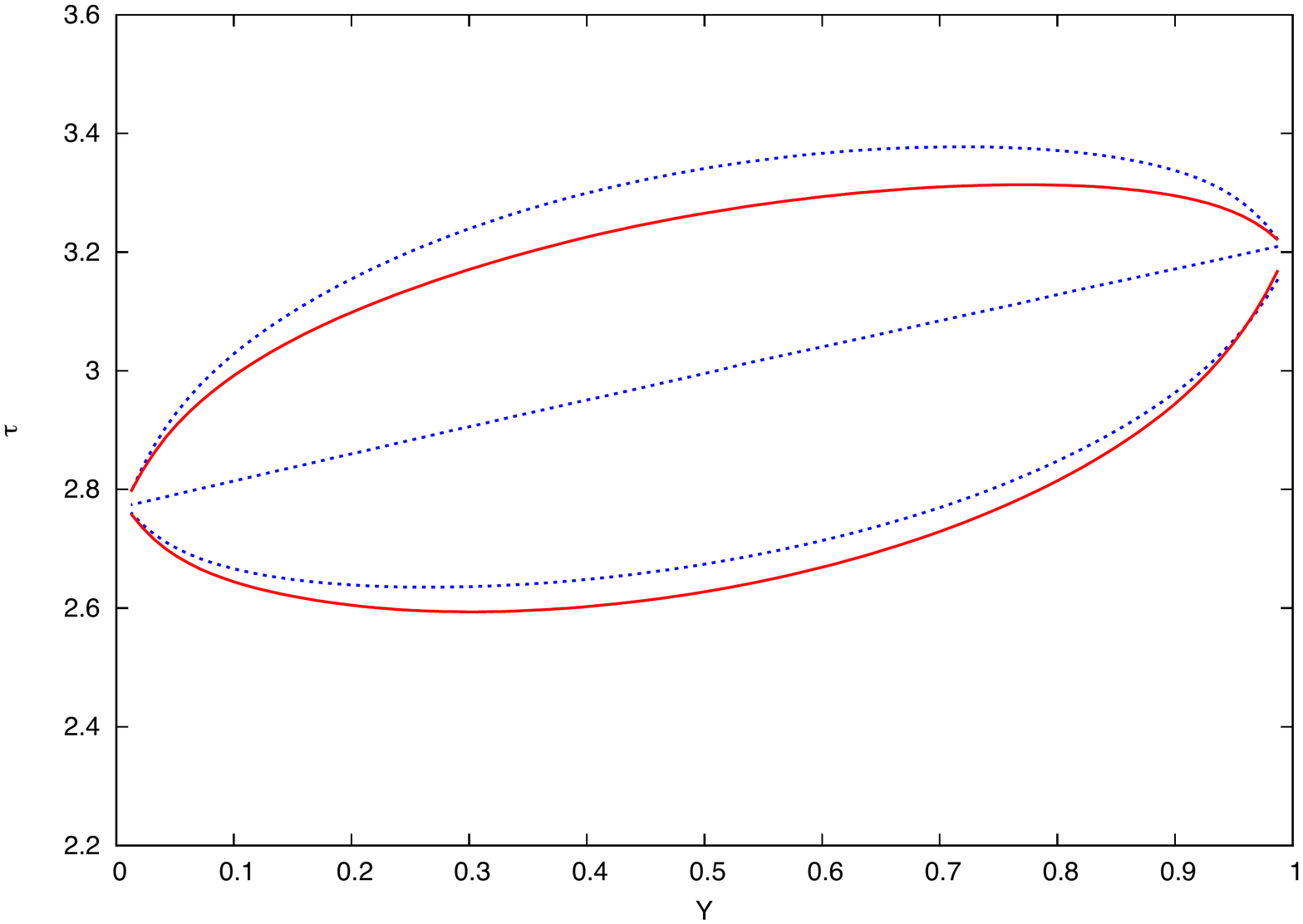}
    \vspace{-1cm}
  \caption{Diffusion pattern for fixed coupling (blue, dotted line) and running coupling (red, solid line).}
  \label{fig:diffusion}
    \vspace{-.3cm}
\end{wrapfigure}

\begin{figure}
\centering
\subfloat[Distribution in the contributions to the BFKL gluon Green function 
    with a fixed number of iterations of the kernel, plotted for different values of the center-of-mass energy, and a fixed ${\bar \alpha}_s = 0.2$. Left: forward case; right: non-forward case with $q=5$ GeV.]{\includegraphics[width=6.5cm,angle=0]{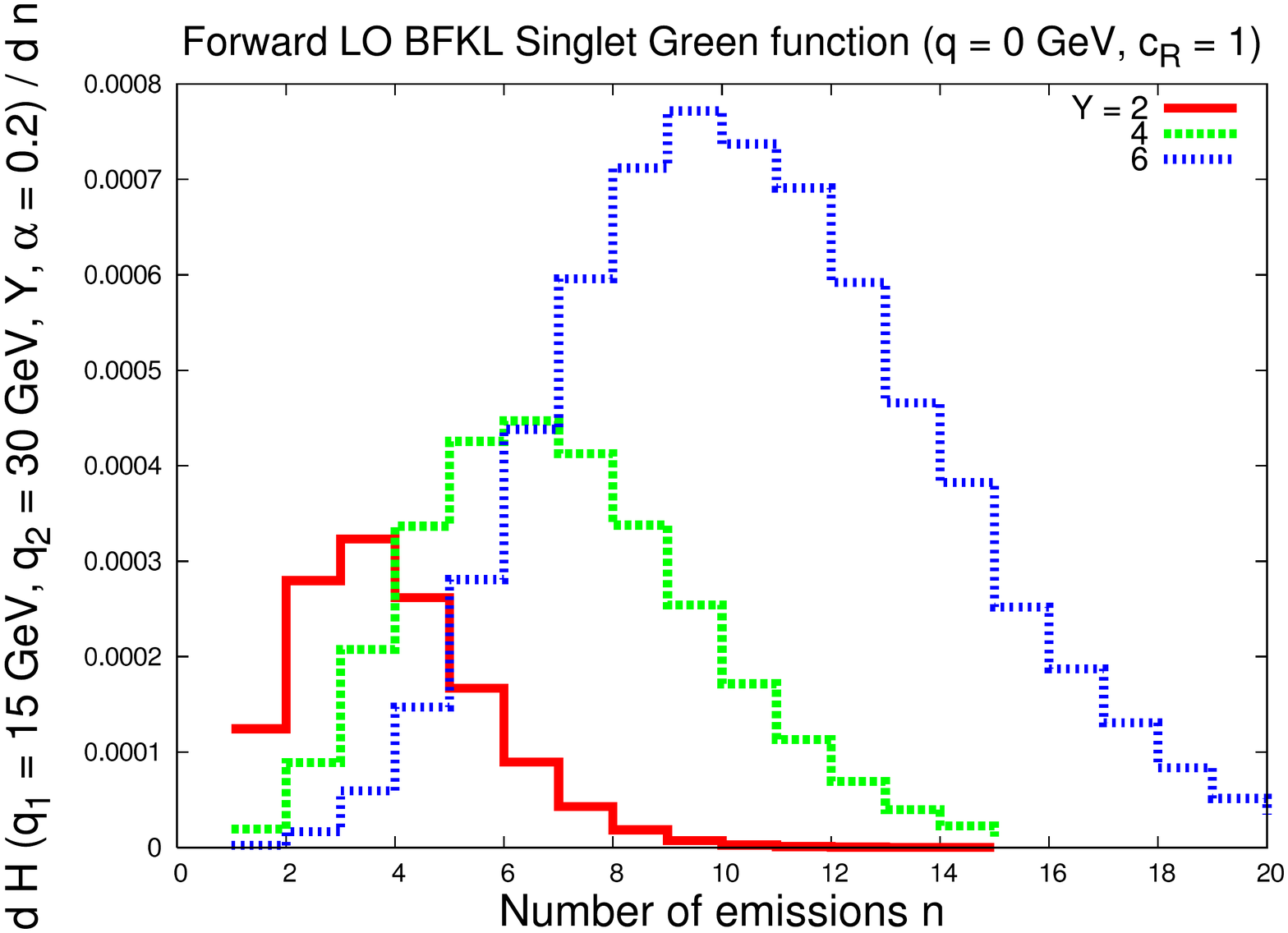}\includegraphics[width=6.5cm,angle=0]{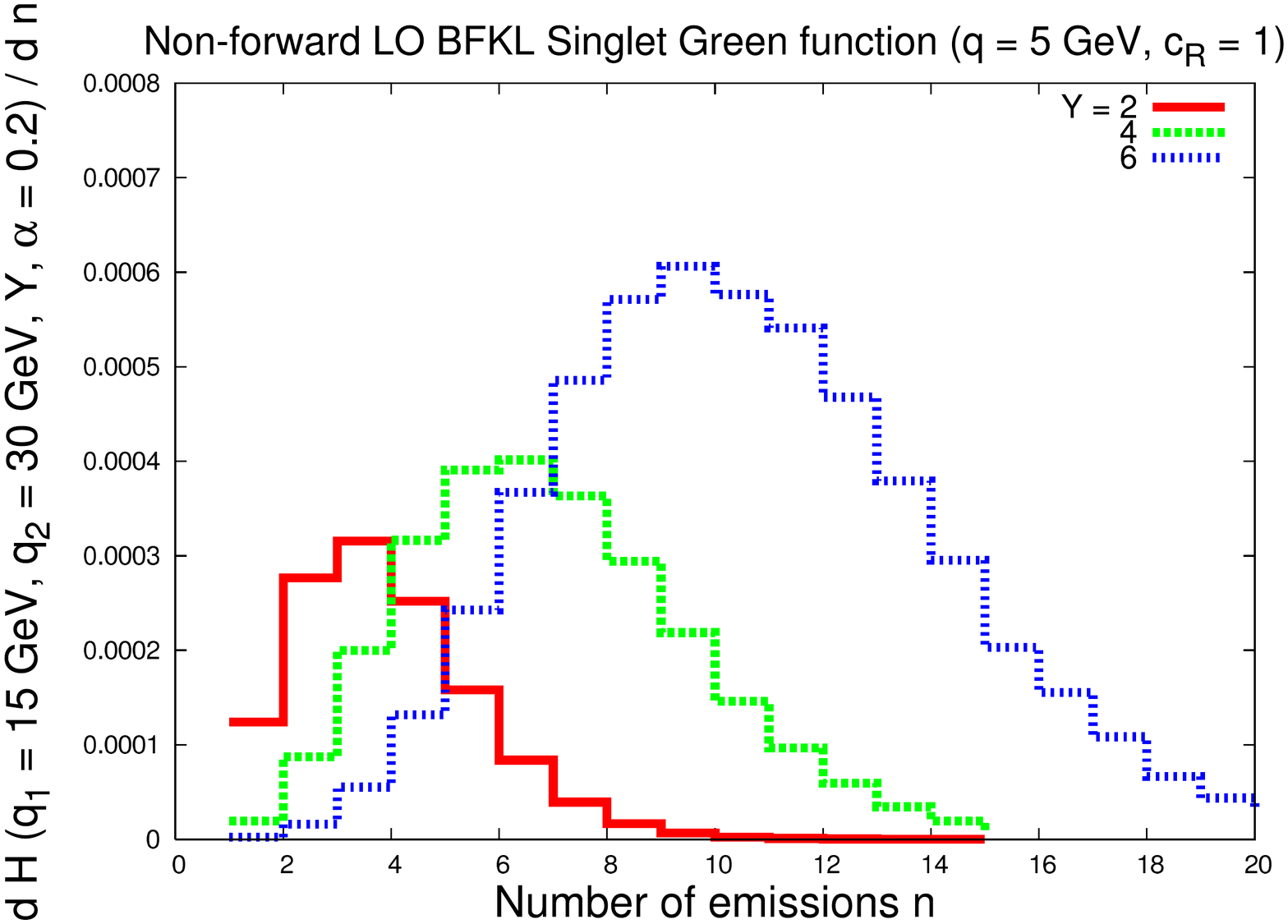} \label{NumberOfTerms-a}} \\
\subfloat[Projection of the gluon Green function on different Fourier components in the azimuthal angle between the ${\bf q}_1$ and ${\bf q}_2$ transverse momenta. Left: forward case; right: non-forward case with $q=5$ GeV.]{\includegraphics[width=6.5cm,angle=0]{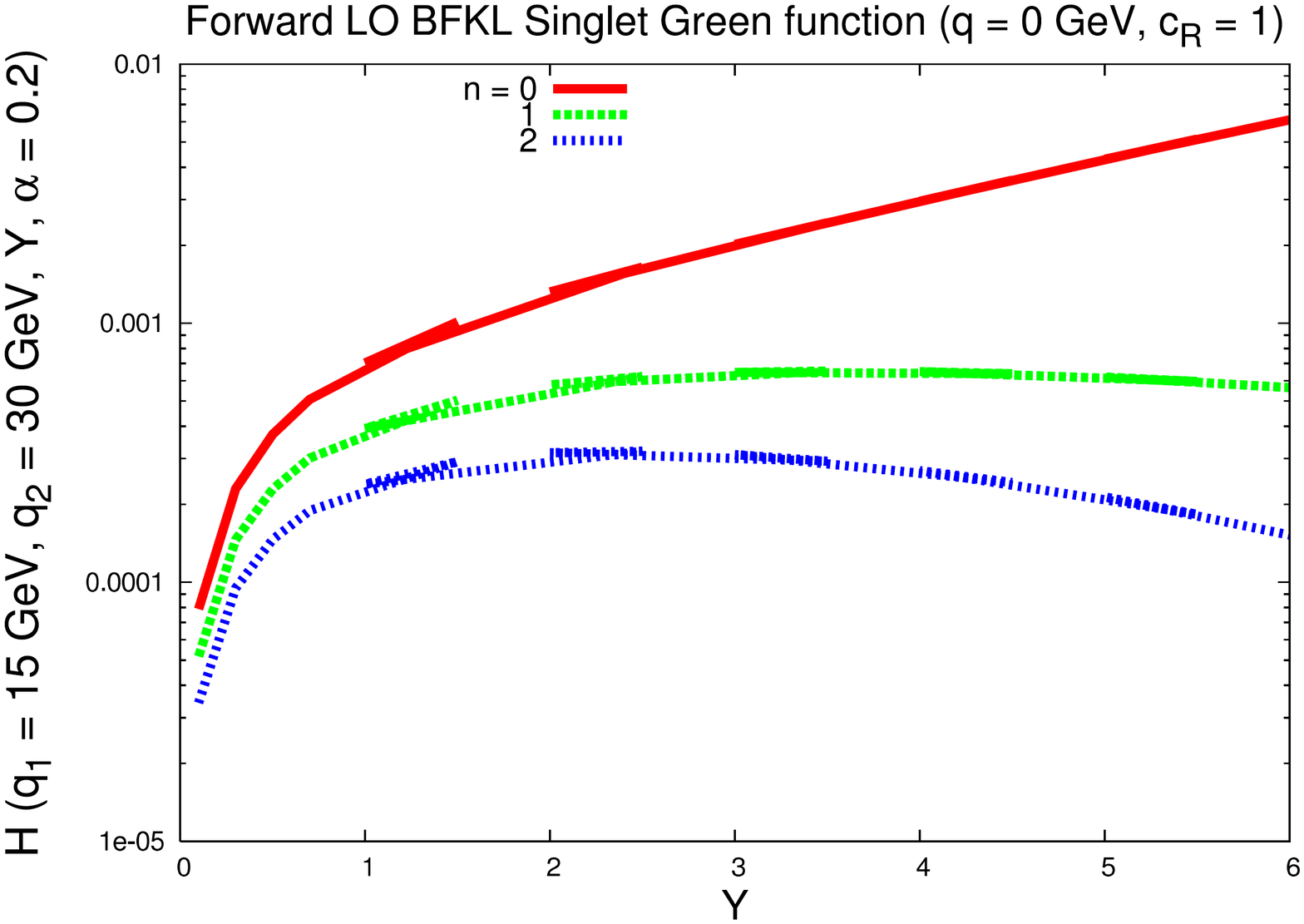} \label{Angles-n-a}\includegraphics[width=6.5cm,angle=0]{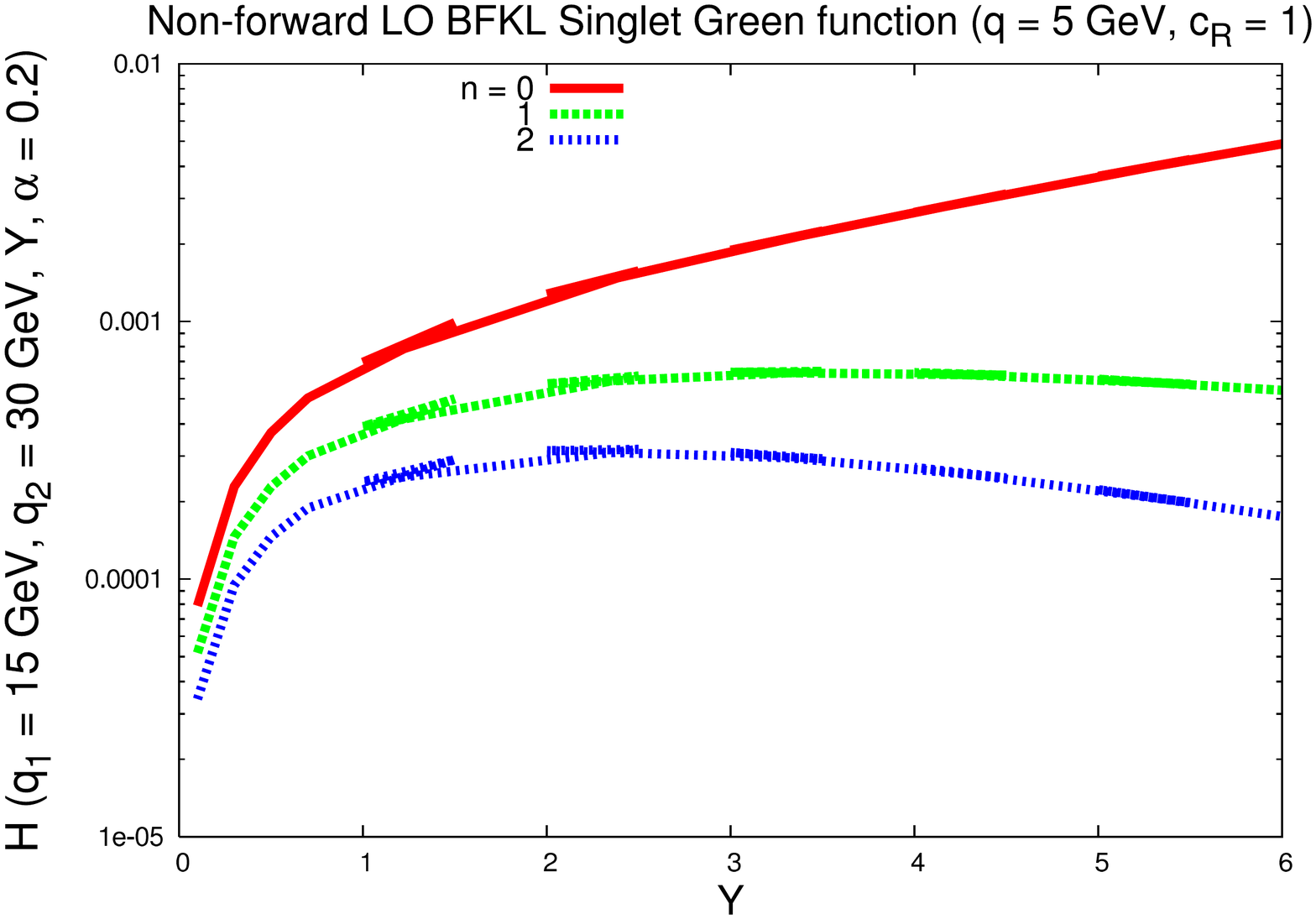}}\\
\subfloat[Gluon Green function dependence for the full range in azimuthal angles.]{\includegraphics[width=6.5cm,angle=0]{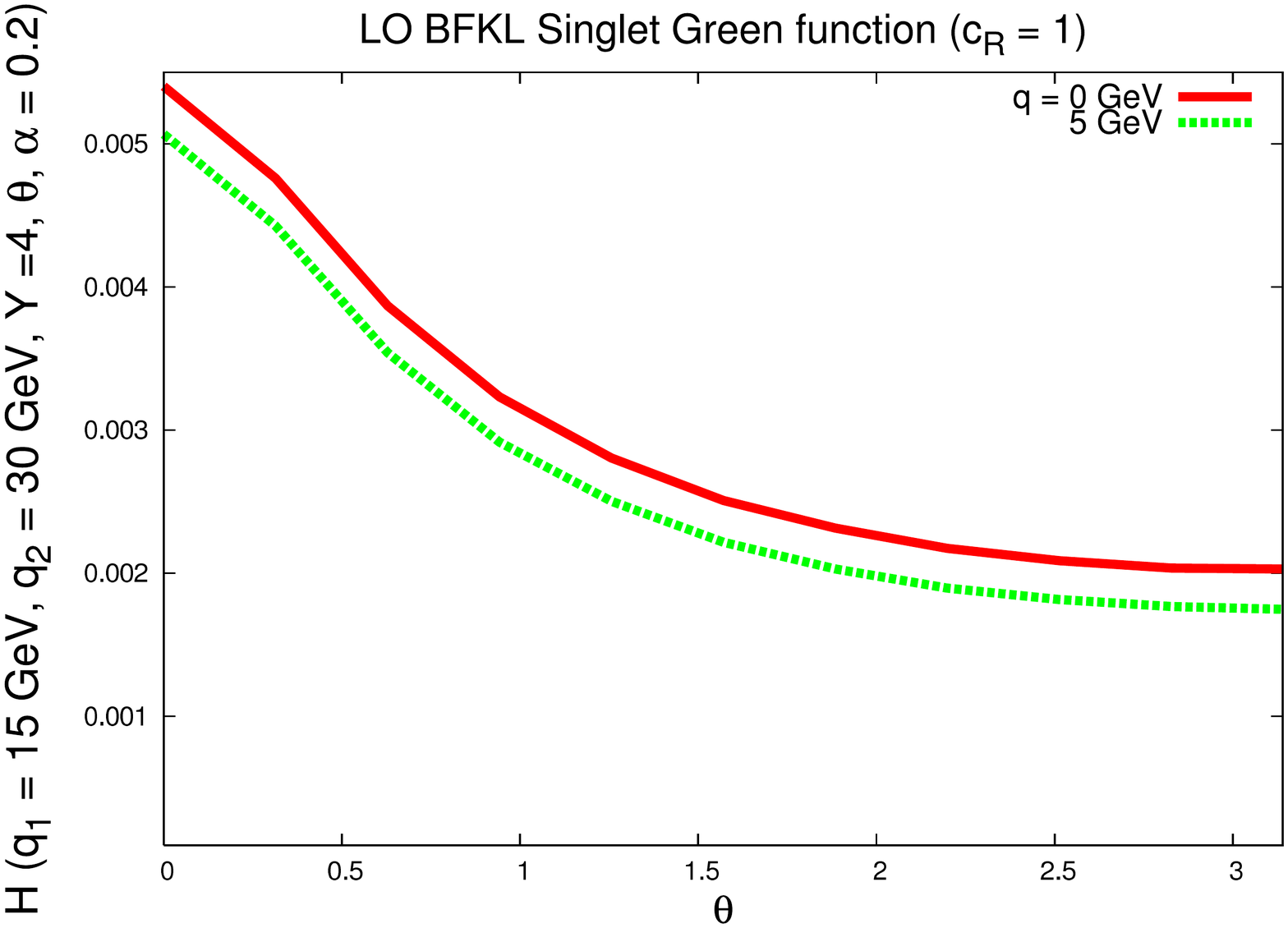} \label{BFKLfullangle-a}}\hspace{.3cm}
\subfloat[Collinear behavior of the gluon Green function.]{\includegraphics[width=6.5cm,angle=0]{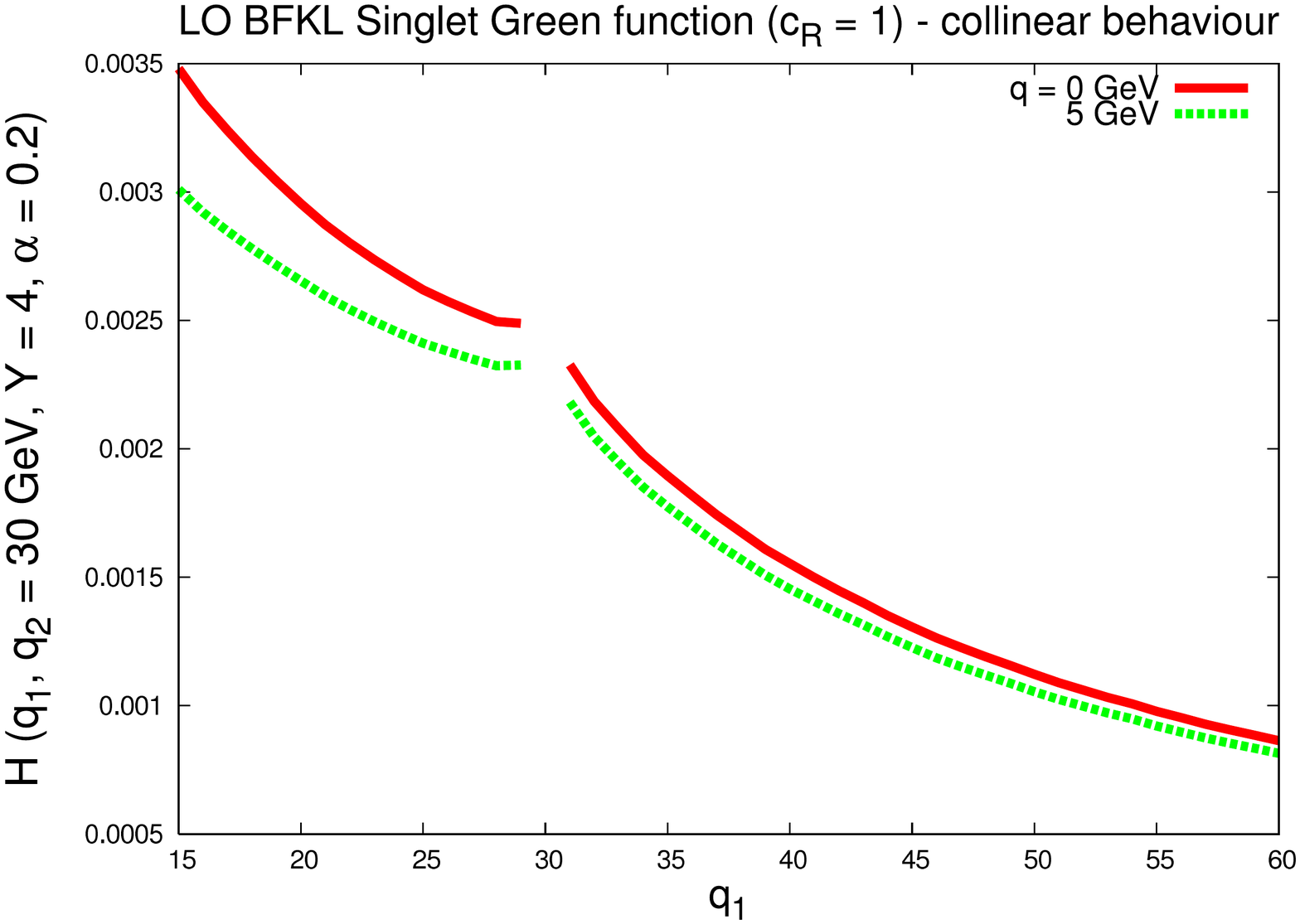} \label{BFKLfullangle-b}}
      \caption{Numerical analysis}
\end{figure}

Figures~\ref{NumberOfTerms-a} show the convergence of the sum defining the function ${\cal H}\left({\bf q}_1,{\bf q}_2;{\bf q};{\rm Y}\right)$. In can be seen how for a fixed value of Y and the coupling ${\bar \alpha}_s$ a finite number of terms in the sum is enough to have a good accuracy for the gluon Green function. As the value of the effective parameter ${\bar \alpha}_s {\rm Y}$ gets larger the Green function is more sensitive to high multiplicity terms, following a Poissonian distribution. It can also be noticed how the distribution in the number of iterations of the kernel gets broader for larger center of mass energies although the convergence is always good. 

It is also instructive to study the solution in terms of the different Fourier 
components in the azimuthal angle between the two momenta ${\bf q}_1$ and 
${\bf q}_2$. A complete analysis is shown in Figs.~\ref{Angles-n-a} for both forward and non-. It can be seen how the only rising component is the $n=0$ one. For completeness, fig.~\ref{BFKLfullangle-a} shows the dependence of the solution on the azimuthal angle for the sum of all Fourier components. The collinear limit is investigated in Fig.~\ref{BFKLfullangle-b}. 

\subsection{Diffusion}\label{sec:diffusion}

The diffusion~\cite{Forshaw:1997dc,Bartels:1995yk} of the transverse scales in the BFKL ladder has been studied in terms of the average value $\langle\tau\rangle$ of $\tau = \log\left(\left(\bq_1+\sum \bk_i\right)^2\right)$ as a function of the rapidity $Y'$ along the gluon ladder. For each set $[\bq_1,\bq_2, Y]$ (where $\bq_1$ and $\bq_2$ are the transverse momenta of the edges of the ladder) the non-forward BFKL equation is solved numerically allowing the study of the evolution of $\langle\tau\rangle$ along the ladder as well as the weight of each configuration point in $n$-momenta phase space to the total solution. Fig.~\ref{fig:diffusion} compares the diffusion pattern found for a fixed choice of the strong coupling and a version with running coupling inserted as explained in the previous section. The straight line in the middle corresponds to $\langle\tau\rangle$ while the upper/lower curves are the mean plus/minus the standard deviation. The set of values taken for the plot are $k_a=5$ GeV, $k_b = 4$ GeV and $Y=1$. The figure shows how the version with running is shifted to the infrared providing a smooth transition from the hard to the soft pomeron.

\section*{Acknowledgements}

The European Comission (LHCPhenoNet PITN-GA-2010-264564) is acknowledged for funding the expenses of the conference.

{\raggedright
\begin{footnotesize}
\bibliographystyle{DISproc}
\bibliography{citations.bib}
\end{footnotesize}
}


\end{document}